\documentclass{emulateapj}
\usepackage{natbib,epsfig}

\begin{document}

\newcommand{\msun}{\ifmmode{M}_{\odot}\else${M}_{\odot}$\fi}
\newcommand{\lsun}{\ifmmode\mbox{L}_{\odot}\else$\mbox{L}_{\odot}$\fi}
\newcommand{\rsun}{\ifmmode\mbox{R}_{\odot}\else$\mbox{R}_{\odot}$\fi}
\newcommand{\degrees}{\ifmmode^{\circ}\else$^{\circ}$\fi}
\newcommand{\amin}{\ifmmode^{\prime}\else$^{\prime}$\fi}
\newcommand{\asec}{\ifmmode^{\prime\prime}\else$^{\prime\prime}$\fi}

\newcommand{\capone}{\caption{}}

\slugcomment{\today, submitted to ApJ}
\shortauthors{Stairs et al.}
\shorttitle{Wide-orbit binary pulsars}

\title{Discovery of Three Wide-orbit Binary Pulsars: Implications for Binary 
Evolution and Equivalence Principles}

\author{I.~H.~Stairs\altaffilmark{1},
A.~J.~Faulkner\altaffilmark{2},
A.~G. Lyne\altaffilmark{2},
M. Kramer\altaffilmark{2},
D.~R.~Lorimer\altaffilmark{2},
M.~A.~McLaughlin\altaffilmark{2},
R.~N. Manchester\altaffilmark{3},
G.~B.~Hobbs\altaffilmark{3},
F. Camilo\altaffilmark{4},
A. Possenti\altaffilmark{5},
M. Burgay\altaffilmark{5},
N. D'Amico\altaffilmark{5,6},
P.~C.~Freire\altaffilmark{7}, and
P.~C.~Gregory\altaffilmark{1}}

\altaffiltext{1}{Dept.~of Physics and Astronomy, University of British
  Columbia, 6224 Agricultural Road, Vancouver, BC V6T~1Z1, Canada; stairs@astro.ubc.ca}
\altaffiltext{2}{Jodrell Bank Observatory, University of Manchester, 
  Macclesfield, Cheshire SK11 9DL, U. K.}
\altaffiltext{3}{Australia Telescope National Facility, CSIRO, P.O.~Box~76,
Epping NSW~1710, Australia}
\altaffiltext{4}{Columbia Astrophysics Laboratory, Columbia University,
550 West 120th Street, New York, NY~10027}
\altaffiltext{5}{INAF - Osservatorio Astronomico di Cagliari, Loc. Poggio dei Pini, Strada 54, 09012 Capoterra (CA), Italy}
\altaffiltext{6}{Universita degli Studi di Cagliari, Dipartimento fi Fisica, SP Monserrato-Sestu km 0.7, 90042, Monserrato (CA), Italy}
\altaffiltext{7}{NAIC, Arecibo Observatory, Barrio Esperanza, Arecibo, PR 00613}

\begin{abstract}
  We report the discovery of three binary millisecond pulsars during
  the Parkes Multibeam Pulsar Survey of the Galactic Plane.  The
  objects are highly recycled and are in orbits of many tens of days
  about low-mass white-dwarf companions.  The eccentricity of one
  object, PSR~J1853+1303, is more than an order of magnitude lower
  than predicted by the theory of convective fluctuations during tidal
  circularization.  We demonstrate that, under the assumption that the
  systems are randomly oriented, current theoretical models of the
  core-mass--orbital-period relation for the progenitors of these
  systems likely overestimate the white-dwarf masses, strengthening
  previous concerns about the match of these models to the data.  The
  new objects allow us to update the limits on violation of
  relativistic equivalence principles to 95\% confidence upper limits
  of $5.6\times 10^{-3}$ for the Strong Equivalence Principle
  parameter $|\Delta|$ and $4.0\times10^{-20}$ for the
  Lorentz-invariance/momentum-conservation parameter $|\hat
  \alpha_3|$.
\end{abstract}


\section{Introduction \label{sec:intro}}

Millisecond radio pulsars are the product of an extended period of
mass and angular momentum transfer to a neutron star (NS) from an
evolving companion star.  This ``recycling'' scenario was proposed
\citep{bk74,sb76,acrs82} shortly after the discovery of the first binary and
recycled pulsars \citep{ht75a,bkh+82}.  Overviews of the mass
transfer process are given in several places
\citep[e.g.,][]{bv91,pk94,tv03}.  It has become clear that there are
in fact several sub-classes of recycled pulsars, the most obvious
distinction being between those that have NS versus white-dwarf (WD)
companions.  Even within the latter category and focusing on Galactic
field binaries only, a wide range of companion masses and evolutionary
histories are represented.  One sub-group is the ``intermediate-mass
binary pulsars'' \citep[e.g.,][]{clm+01,eb01b} which have spin periods
of tens of milliseconds and/or companions that are likely massive (CO
or ONeMg) WDs.  Many of these systems may have experienced an
ultra-high mass transfer rate, or else undergone a period of
common-envelope evolution \citep{vdh94,tkr00,tvs00}.  Among pulsars
with lower-mass Helium WD companions, there is another split.  Systems
whose initial orbital periods were less than the ``bifurcation''
period of about 1 or 2 days \citep{ps88,esa98} will see their orbits
shrink through magnetic braking and gravitational radiation,
ultimately forming a low-mass binary pulsar in a tight orbit.  Systems
with longer initial orbital periods undergo stable, long-lived mass
transfer via an accretion disk, once the companion star has evolved
onto the giant branch.  The resulting systems have long orbital
periods of several days or more.  The prototype of this ``wide-orbit
binary millisecond pulsar'' (WBMSP) group, PSR~B1953+29, was one of
the first recycled pulsars discovered \citep{bbf83}; over the years
the number of these pulsars with orbital periods greater than 4 days
has grown to 18.

The WBMSPs are the best-understood class of pulsar--WD binaries.  For
instance, the companion mass and orbital separation are thought to
follow the ``core-mass--orbital-period'' ($P_{\rm b}$--$m_2$)
relation, in which the mass of the core which eventually forms the
white dwarf is directly related to the size of the envelope of the
Roche-Lobe-filling giant star and hence the orbital radius
\citep[e.g.,][]{rpj+95,ts99a}.  Measurements of companion masses to date
(via pulsar timing or optical spectroscopy) indicate that this
relation is fairly well satisfied
\citep{ktr94,vbb+01,sns+05,vbjj05}, although we shall discuss this 
issue further below.  Furthermore, although the orbit becomes tidally
circularized during the companion's giant phase, fluctuations of
convective cells in the giant's envelope force the eccentricity to a
non-zero value \citep{phi92b}; thus, to within an order of magnitude
or so, the eccentricities of these WBMSP systems can be predicted from
the orbital periods.

More examples of WBMSP pulsars are needed to further test these
predictions, and also to provide additional constraints for population
synthesis efforts \citep[e.g.,][]{py98,wk02,wk03a,prp03}.  These
objects are also valuable for constraining departures from general
relativity (GR) in the form of equivalence principle violations
\citep[e.g.,][]{ds91}.  Finally, some pulsar-WD binaries permit
measurement of NS and WD masses through relativistic or geometric
timing effects; thus new systems potentially add to the pool of
objects that can be used to constrain theories of the NS interior
and/or the amount of matter transferred in different evolutionary
processes \citep[e.g.,][]{sta04c}.  In this paper we report the
discovery of three more WBMSPs during the Parkes Multibeam Pulsar
Survey.  In \S~\ref{sec:obs} we describe the search and follow-up
timing observations.  In \S~\ref{sec:psrs} we describe the
characteristics of the new pulsars and relate them to evolutionary
theory including a comparison with the predictions of the $P_{\rm
b}$--$m_2$ relation.  In \S~\ref{sec:gr} we use the ensemble of
pulsars with white-dwarf companions to set new, stringent limits on
equivalence principle violations.  Finally, in
\S~\ref{sec:conc} we summarize our results and look to the
future.
                                                                               
\section{Observations and Data Analysis}
\label{sec:obs}

The Parkes Multibeam Pulsar Survey \citep[e.g.,][]{mlc+01} used a 13-beam
receiver on the 64-m Parkes telescope to search the Galactic Plane
($|b|<5^{\circ}$, $260^{\circ}<l<50^{\circ}$) for young and recycled
pulsars.  Observations were carried out at 1374\,MHz with 96 channels
across a 288\,MHz bandpass, using 35-minute integrations and 0.25\,ms
sampling.  More than 700 pulsars have been discovered
\citep[e.g.,][]{hfs+04}, nearly doubling the previously known population.  The
survey processing includes dedispersion of the data at numerous trial
dispersion measures, followed by a periodicity search using Fast
Fourier Transforms and harmonic summing \citep{mlc+01}.  Recently, the
entire data volume has been reprocessed using ``acceleration''
searches for pulsars in fast binary orbits, as well as improved
interference excision techniques.  This has resulted in a number of
new binary and millisecond pulsars (Faulkner et al. 2004, 2005),
\nocite{fsk+04,fkl+05} including two of the objects described in this
paper.

PSR~J1751$-$2857 has been observed since MJD 51972 with Parkes at
1390\,MHz using a 512-channel filterbank across 256\,MHz bandwidth and
0.25\,ms sampling and with the 76\,m Lovell Telescope at Jodrell Bank
Observatory at 1396\,MHz using a 64-channel filterbank across 64 MHz
with 0.13\,ms sampling.

PSRs~J1853+1303 and J1910+1256 were discovered in observations taken
on MJDs 52321 and 52322, respectively, and identified as candidates
using the {\sc reaper} program selection procedure \citep{fsk+04}.
Both were confirmed as pulsars on MJD 52601.  Subsequently, they have
been timed at Parkes and Jodrell Bank using the systems described
above.  Both have also been regularly observed using the 305\,m
Arecibo telescope in Puerto Rico.  These observations initially used
one 100\,MHz Wideband Arecibo Pulsar Processor (WAPP) in a
fast-sampled ``search'' mode, and, since 2004 Feb., 3 WAPPs in an
online folding mode, in all cases with 256 lags and 64\,$\mu$sec
sampling.  The single-WAPP observations were centered at 1400\,MHz and
when 3 WAPPs were used they were centered at frequencies of 1170, 1370
and 1470\,MHz.  Some of the single-WAPP observations display timing
systematics similar in both pulsars relative to the contemporaneous
Parkes data; as this likely reflects instrumental errors in the
fairly new WAPP, these data points have been left out of the timing
analysis.  Some of the Arecibo observations were flux calibrated using
a pulsed noise diode of known strength; the resulting calibrated
profiles are used to determine the flux densities for PSRs~J1853+1303
and J1910+1256.  The flux density for PSR~J1751$-$2857 was determined
using the procedure outlined in \citet{hfs+04}.

\begin{figure}
  \plotone{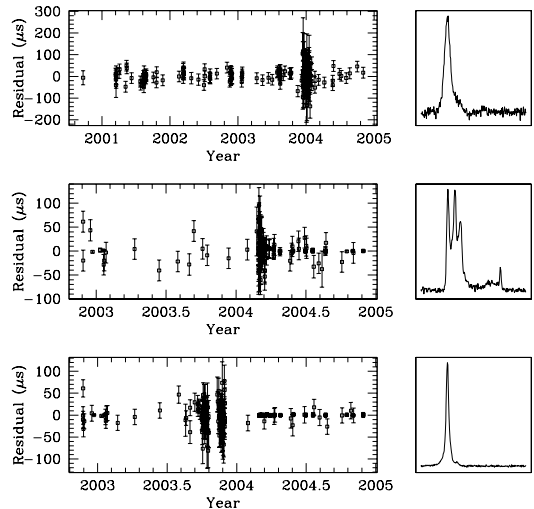} \caption{Timing residuals and full-period profiles
  at 1400 MHz for each of the three new pulsars. Top:
  PSR~J1751$-$2857.  Middle: PSR~J1853+1303.  Bottom: PSR~J1910+1256.
  The profile for PSR~J1751$-$2857 was obtained with Parkes; the other
  two with Arecibo.  For each pulsar, the TOAs with large scatter are
  from Jodrell Bank for all pulsars and from Arecibo for
  PSR~J1853+1303, and represent eras when the timing solutions were
  poorly known.} \label{fig:resids}
\end{figure}

The data from each telescope were dedispersed and folded modulo the
predicted topocentric pulse period; this was accomplished off-line for
the Parkes data and Arecibo search-mode data and on-line for the
Jodrell Bank and Arecibo folding-mode data.  A Time-of-Arrival (TOA)
was determined for each observation by cross-correlation with a high
signal-to-noise standard template \citep{tay92}.  The timestamp for
each observation was based on the observatory time standard, corrected
by GPS to Universal Coordinated Time (UTC).  The JPL DE200 ephemeris
\citep{sta90} was used for barycentric corrections.  The timing
solutions were found using the standard pulsar timing program
{\sc tempo}\footnote{{\tt http://pulsar.princeton.edu/tempo}}, with
uncertainties containing a small telescope-dependent amount added in
quadrature and scaled to ensure $\chi^2_{\nu} \simeq 1$.  The
resulting pulsar parameters are shown in Table~\ref{tab:params}, while
the timing residuals and standard pulse profiles are shown in
Figure~\ref{fig:resids}.  We note that the correctly folded Arecibo
data have rms residuals on the order of 1\,$\mu$sec per WAPP for
PSR~J1910+1256 and 1--2\,$\mu$sec per WAPP for PSR~J1853+1303
for roughly 30-minute integrations.

\begin{deluxetable*}{lrrr}[t]
\footnotesize
\tablecaption{Parameters for the new pulsars \label{tab:params}}
\tablewidth{0pt}
\tablehead{\colhead{Parameter} & \colhead{J1751$-$2857}& \colhead{J1853+1303} & \colhead{J1910+1256}}
\startdata
Right ascension, $\alpha$ (J2000) \dotfill  & $17^{\rm h}\,51^{\rm m}\,32\fs6965(2)$ & $18^{\rm h}\,53^{\rm m}\,57\fs31827(8)$ & $19^{\rm h}\,10^{\rm m}\,09\fs70041(6)$ \\
Declination,     $\delta$ (J2000) \dotfill  & $-28\degrees\,57\amin\,46\farcs50(3)$  & $13\degrees\,03\amin\,44\farcs0884(17)$  & $12\degrees\,56\amin\,25\farcs5276(6)$ \\
Pulse period, $P$ (ms)           \dotfill   & 3.9148731963690(6)  & 4.0917973806819(14)  & 4.9835839397055(12) \\
Period derivative, $\dot P$ (s\,s$^{-1}$)  \dotfill & 1.126(4)$\times$10$^{-20}$ & 8.85(10)$\times$10$^{-21}$ & 9.77(7)$\times$10$^{-21}$ \\
Epoch (MJD)                    \dotfill             & 52560.0 & 52972.0 & 52970.0 \\
Dispersion measure (pc\,cm$^{-3}$)  \dotfill         & 42.808(20) & 30.5702(12) & 38.0650(7) \\
Orbital period, $P_{\rm b}$ (d)    \dotfill        & 110.7464576(10) & 115.6537868(4) & 58.46674201(9) \\
Projected semi-major axis, $x$ (lt-s)\dotfill & 32.528221(9) & 40.7695200(10) & 21.1291045(6) \\
Eccentricity, $e$                  \dotfill         &  0.0001283(5) &  0.00002369(9) &  0.00023022(6) \\
Longitude of periastron, $\omega$ (deg) \dotfill  & 45.52(19)\tablenotemark{a}  & 346.63(9)\tablenotemark{a} & 106.001(11)\tablenotemark{a} \\
Epoch of periastron, $T_0$ (MJD) \dotfill  & 52491.58(6)\tablenotemark{a}  & 52890.25(3)\tablenotemark{a} & 52968.4474(18)\tablenotemark{a} \\
Data span (MJD)          \dotfill         & 51808--53312  & 52606--53337 & 52602--53337\\
Number of TOAs                     \dotfill         & 168 & 140 & 183 \\
Weighted RMS timing residual ($\mu$s)   \dotfill    & 28.5 & 2.9 & 1.8 \\
Flux Density at 1400\,MHz, $S_{1400}$ (mJy)\dotfill & 0.06(2) & 0.4(2) & 0.5(1) \\
\cutinhead{Derived Parameters}
Galactic longitude, $l$ (deg) \dotfill  & 0.65 & 44.87 & 46.56 \\
Galactic latitude, $b$ (deg) \dotfill  & $-1.12$ & 5.37 & 1.80 \\
Distance \citep{cl02} (kpc) \dotfill & 1.4 & 1.6 & 1.9 \\
Mass function, $f_1$ (\msun)     \dotfill    & 0.003013034(2) & 0.0054396358(4) & 0.0029628402(2)  \\
Minimum companion mass, $m_2$ (\msun)\tablenotemark{b}   \dotfill    & 0.19 & 0.24 & 0.22 \\
Surface magnetic field, $B = 3.2 \times 10^{19}\sqrt{P \dot P}$ (G) \dotfill & $2.1\times 10^8$ & $1.9\times 10^8$ & $2.2\times 10^8$ \\
Characteristic age, $\tau_c = P/{2 \dot P}$ (Gyr) \dotfill & 5.5 & 7.3 & 8.1 \\
\enddata
\tablecomments{Figures in parentheses are uncertainties in the last digits
quoted, which are twice the formal errors reported by {\sc tempo} after scaling the TOA uncertainties to obtain a reduced-$\chi^2$ of about 1.0.}
\tablenotetext{a}{The parameters $\omega$ and $T_0$ are highly covariant.  Observers should use 
for J1751$-$2857: $\omega$ = 45.523832$^{\circ}$, $T_0$ = 52491.578696239; 
for J1853+1303: $\omega$ = 346.630447$^{\circ}$, $T_0$ = 52890.248760182; 
for J1910+1256: $\omega$ = 106.001079$^{\circ}$, $T_0$ = 52968.447431428.}
\tablenotetext{b}{Assuming a pulsar mass of 1.35\,\msun.}
\end{deluxetable*}

\section{Discussion \label{sec:discussion}}
\subsection{Pulsar Parameters and Evolution \label{sec:psrs}}

\begin{deluxetable*}{lrrlrrrrc}[t]
\footnotesize
\tablecaption{The 21 pulsars with low-mass WD companions and orbital periods of more than 4 days \label{tab:allbin}}
\tablewidth{0pt}
\tablehead{\colhead{Pulsar} & \colhead{$P$ (ms)}& \colhead{$P_{\rm b}$ (days)} & \colhead{$e$} & \colhead{$\omega$($^{\circ}$)} & \colhead{PM RA} & \colhead{PM Dec} & \colhead{$f_1$} & \colhead{References}\\
\colhead{} & \colhead{}& \colhead{} & \colhead{} & \colhead{} & \colhead{(mas/yr)} & \colhead{(mas/yr)} & \colhead{(\msun)} & \colhead{}}
\startdata
J0407+1607 & 25.702 & 669.0704 & 0.0009368(6) & 192.74(2) & -- & -- &
0.002893 & 1 \\ 
J0437$-$4715 & 5.757 & 5.7410 &
0.0000191686(5) & 1.20(5) & 121.438(6) & $-$71.438(7) & 0.001243 & 2,3\\ 
J1045$-$4509 & 7.474 &
4.0835 & 0.0000197(13) & 243(4) & $-$5(2) & 6(1) & 0.001765 & 4,5 \\ 
J1455$-$3330 & 7.987 & 76.1746 &
0.0001697(3) & 223.8(1) & 5(6) & 24(12) & 0.006272 & 6,5 \\ 
J1640+2224 & 3.163 & 175.4606 &
0.000797262(14) & 50.7308(10) & 1.66(12) & $-$11.3(2) & 0.005907 & 7,8\\ 
J1643$-$1224 & 4.622 &
147.0174 & 0.0005058(1) & 321.81(1) & 3(1) & $-$8(5) & 0.000783 & 6,5\\ 
J1709+2313 & 4.631 & 22.7119
& 0.0000187(2) & 24.3(6) & 3.2(7) & 9.7(9) & 0.007438 & 7,9 \\ 
J1713+0747 & 4.570 & 67.8255 & 0.0000749406(13) &
176.1915(10) & 4.917(4) & $-$3.933(10) & 0.007896 & 10,11 \\ 
J1732$-$5049 & 5.313 & 5.2630 &
0.0000098(20) & 287(12) & -- & -- & 0.002449 & 12 \\ 
J1751$-$2857 & 3.915 & 110.7465 & 0.0001283(5) &
45.52(19) & -- & -- & 0.003013 & This work \\ 
J1804$-$2717 & 9.343 &
11.1287 & 0.000035(3) & 160(4) & -- & -- & 0.003347 & \ 13 \\ 
J1853+1303 & 4.092 & 115.6538 & 0.00002369(9) &
346.63(8) & -- & -- & 0.005440 & This work \\ 
B1855+09 & 5.362 &
12.3272 & 0.00002170(3) & 276.39(4) & $-$2.899(13) & $-$5.45(2) &
0.005557 & 14,15 \\ 
J1910+1256
& 4.984 & 58.4667 & 0.00023022(6) & 106.001(11) & -- & -- & 0.002963 &
This work \\ 
J1918$-$0642 & 7.646 & 10.9132 & 0.000022(4) & 234(11) &
-- & -- & 0.005249 & 12 \\ 
B1953+29 & 6.133 &
117.3491 & 0.0003303(1) & 29.55(2) & $-$1.0(3) & $-$3.7(3) & 0.002417
& 16,17 \\ 
J2016+1948 & 64.940
& 635.039 & 0.00128(16) & 90(5) & -- & -- & 0.009112 & \ 18 \\ 
J2019+2425 & 3.935 & 76.5116 & 0.00011109(4) &
159.03(2) & $-$9.41(12) & $-$20.60(15) & 0.010687 & 19,20\\ 
J2033+1734 & 5.949 & 56.3078 &
0.00012876(6) & 78.23(3) & $-$5.94(17) & $-$11.0(3) & 0.002776 & 21,15\\ 
J2129$-$5721 & 3.726 &
6.6255 & 0.0000068(22) & 178(12) & 7(2) & $-$4(3) & 0.001049 & 13,5 \\
J2229+2643 & 2.978 & 93.0159 &
0.0002556(2) & 14.42(5) & 1(4) & $-$17(4) & 0.000839 & 22,17 \\ 
\enddata
\tablerefs{ 1. \citet{lxf+05}\label{tref:lor}, 2. \citet{jlh+93}\label{tref:jlh93}, 3. \citet{vbb+01}\label{tref:vbb01}, 4. \citet{bhl+94}\label{tref:bhl94}, 5. \citet{tsb+99}\label{tref:tsb99}, 6. \citet{lnl+95}\label{tref:lnl95}, 7. \citet{fcwa95}\label{tref:fcwa95}, 8. \citet{llww05}\label{tref:llww05}, 9. \citet{lwf+04}\label{tref:lwf+04}, 10. \citet{fwc93}\label{tref:fwc93}, 11. \citet{sns+05}\label{tref:sns05}, 12. \citet{eb01b}\label{tref:eb01b}, 13. \citet{llb+96}\label{tref:llb96}, 14. \citet{srs+86}\label{tref:srs86}, 15. \citet{spl04}\label{tref:spl04}, 16. \citet{bbf83}\label{tref:bbf83}, 17. \citet{wdk+00}\label{tref:wdk+00}, 18. \citet{naf03}\label{tref:naf03}, 19. \citet{ntf93}\label{tref:ntf93}, 20. \citet{nss01}\label{tref:nss01}, 21. \citet{rtj+96}\label{tref:rtj96}, 22. \citet{cnt96}\label{tref:cnt96} }
\end{deluxetable*}

\begin{figure}
  \plotone{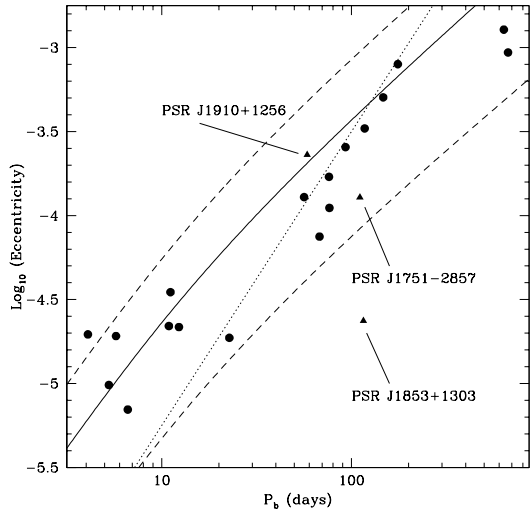} \caption{Eccentricity $e$ vs. orbital period
  $P_{\rm b}$ for the pulsar-WD systems thought to be described by
  stable mass transfer and the $P_{\rm b}$--$m_2$ relation.  The three
  new pulsars are labeled and indicated by triangles.  The solid and
  dashed curves illustrate the eccentricity ranges predicted by
  \citet{phi92b} as a function of orbital period; 95\% of pulsars
  should fall within this range and 90\% of the observed systems do.
  The dotted line has $P_{\rm b}^2 \propto e$, indicating the
  figure-of-merit for tests of the Strong Equivalence Principle
  (SEP). \label{fig:pbe}}
\end{figure}

Each of the new pulsars is in an orbit of several tens of days with a
companion of minimum mass of 0.2 to 0.25\,$M_{\odot}$.  The
eccentricities of PSRs~J1751$-$2857 and J1910+1256 agree well with the
predictions of \citet{phi92b}; however, that of PSR~J1853+1303 is
lower than the predictions by more than an order of magnitude
(Fig.~\ref{fig:pbe}).  As it has a low-mass companion and its spin
period indicates that it is highly recycled, there is little other
reason to believe that its evolution proceeded in any unusual fashion,
so its low eccentricity may simply reflect the natural scatter in the
population.

\begin{figure}
  \plotone{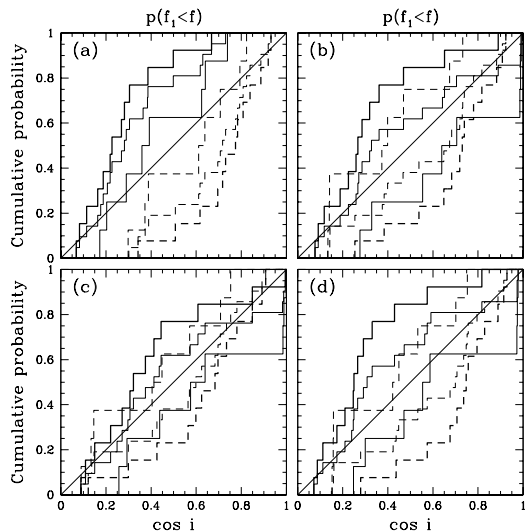} 
\caption{Cumulative probability distributions for the measured mass 
functions $p(f_1<f)$ (solid lines) and the median predicted values of
$\cos i$ (dashed lines) for the 21 binary systems listed in
Table~\ref{tab:allbin}, after \citet{tc99}, Fig. 4.  In each panel,
the thinnest lines incorporate all systems, the medium-weight lines
only those systems with $P_{\rm b} < 50$\,days, and the heaviest lines
only those systems with $P_{\rm b} > 50$\,days.  Panel (a) assumes
$m_2$ is drawn uniformly from the range predicted by the \citet{ts99a}
$P_{\rm b}$--$m_2$ relation and that $m_1$ is drawn from a Gaussian
distribution centered on 1.35\msun with width 0.04\msun \citep{tc99}.
Panel (b) draws $m_1$ from the same range, but assumes $m_2$ is drawn
from the range predicted by the \citet{rpj+95} relation.  Panel (c) as
(a), but drawing $m_1$ from a Gaussian distribution centered on
1.75\msun with width 0.04\msun.  Panel (d) as (b), but with $m_2$
limits given by the \citet{ts99a} fits to the \citet{rpj+95} results.
The straight lines indicate the cumulative probability for a uniform
distribution.}
\label{fig:pbm2}
\end{figure}

There are now 21 pulsars with low-mass WD companions whose orbital
characteristics should be determined by the $P_{\rm b}$--$m_2$
relation (Table~\ref{tab:allbin}).  To date, however, there are only a
handful of observational tests of this relation: three Shapiro-delay
timing measurements and one optical WD spectrum for systems with
orbital periods greater than 2 days \citep{sns+05,vbjj05}.  In order
to judge the agreement of the whole population with the theory,
therefore, we need to use statistical arguments based on the observed
mass functions.  We follow
\citet{tc99} in considering only those pulsars with orbital periods
greater than about 4 days, excluding those in the 2--4 day range as
being too close to the limits of applicability of the relation.  With
this many pulsars, it becomes possible to examine whether different
subgroups are equally well described by the relation.  To do this, we
adopt the approach of
\citet{tc99} (Fig.~4 and related discussion) by assuming a range of
pulsar masses (e.g., 1.35$\pm$0.04\,$M_{\odot}$, which was a good
match to the set of pulsar masses measured at the time of
\citet{tc99}) and the (uniform) $P_{\rm b}$--$m_2$ relation specified
by either \citet{ts99a} or \citet{rpj+95}. For the first test, for
each pulsar we assume a uniform distribution in $\cos i$ and simulate
a large number of systems, finding the probability $p(f_1<f)$ that the
simulated mass function $f$ is above the observed value $f_1$.  For
the second test, we assume the observed $f_1$ and find the median
predicted value of $\cos i$, again simulating a large number of
systems.  The cumulative probability distributions for both $\cos i$
(dashed lines) and $p(f_1<f)$ (solid lines) are displayed in
Fig.~\ref{fig:pbm2}, for all 21 pulsars, and for those with orbital
periods less than (8 systems) and greater than (13 systems) 50 days.
For comparison, the straight lines indicate the cumulative probability
for a uniform distribution.

The $m_2$ limits from the \citet{ts99a} relation (panel (a) of
Fig.~\ref{fig:pbm2}) are given by the Pop. I and Pop. II fits to their
simulation results; they find a spread in $P_{\rm b}$ of a factor of
about 1.4 around their median value for a given $m_2$.  \citet{rpj+95}
find a spread of about 2.4 in $P_{\rm b}$ for any given $m_2$, but
consider this to cover roughly the full range of possible values; we
assume a range of $\sqrt{2.4} \simeq 1.6$ will be comparable to the
\citet{ts99a} ranges and show the corresponding results in panel (b)
of Fig.~\ref{fig:pbm2}.  \citet{ts99a} also provide their own fits
to the \citet{rpj+95} simulations, and we evaluate these fits in panel
(d) of Fig.~\ref{fig:pbm2}.  It is important to note that these
\citet{ts99a} fits do not cover the full spread of the \citet{rpj+95}
orbital periods, favoring the lower periods at any given $m_2$.  Thus
it is perhaps not surprising that panel (d) indicates higher mass
estimates in general than panel (b).

We find that the \citet{ts99a} $P_{\rm b}$--$m_2$ relation is
incompatible at the 99.5\% level (according to a KS-test) with a
uniform distribution of $\cos i$ if the pulsar masses are drawn from a
Gaussian distribution centered on 1.35\msun with width 0.04\msun.
Better agreement with uniformity in $\cos i$ (at the 50\% level) can
be reached if the pulsar masses are very large on average (e.g.,
1.75$\pm$0.04\,$M_{\odot}$; panel (c) of Fig.~\ref{fig:pbm2}), a
situation not supported by observational evidence.  The \citet{rpj+95}
relation appears to be in slightly better agreement with uniformity in
$\cos i$, though it is clear from Fig.~\ref{fig:pbm2} that this occurs
because of a tendency to underestimate the companion masses for
short-period systems and overestimate those for long-period systems.
We note that, although \citet{tc99} favor using the \citet{rj97}
version of the $P_{\rm b}$--$m_2$ relation for $m_2 <
0.25\,M_{\odot}$, this predicts extremely low masses ($\sim 0.10
\,M_{\odot}$) for the shortest-$P_{\rm b}$ systems, which is in
conflict with the observed companion mass of
$0.236\pm0.017\,M_{\odot}$ for PSR~J0437$-$4715
\citep{vbb+01}.  Using this revised relation would lower the estimates of the
companion masses in the short-$P_{\rm b}$ systems even further; this
in combination with the higher estimates for the long-$P_{\rm b}$
systems appears to have been responsible for the overall good
agreement that \citet{tc99} found for the combined relation with NS
masses of 1.35$\pm$0.04\,$M_{\odot}$ and a uniform distribution in
$\cos i$.  Thus, assuming that the cosines of the system inclination
angles are in fact uniformly distributed and that most NS masses are
near 1.35\,$M_{\odot}$, it appears that the existing forms of the
$P_{\rm b}$--$m_2$ relation tend to overestimate companion masses for
long-period systems, while providing conflicting results for the
short-period systems.

The tendency to underestimate masses in the short-$P_{\rm b}$ case was
in fact noted by \citet{rpj+95}, although they included in their
analysis systems now considered to be ``intermediate-mass'' binaries
(such as PSR~J2145$-$0750) having different evolutionary histories.
\citet{tau96} and \citet{ts99a} comment on the poor match of the 
(then 5) known WBMSPs to the higher theoretical predictions of $m_2$.
With the larger number of systems now known, the conclusion of a poor
match seems inescapable.  \citet{rpj+95} note that while the
relationship between core mass and luminosity for red giants is well
understood, the relation between mass or luminosity 
and radius is looser \citep[see also][]{tc99},
with uncertainties in the companion's initial chemical composition and 
the convective mixing-length parameter; this may explain our results.
It appears more theoretical work will be required to derive models that
better match the data.

The only long-orbital-period system with a timing test of the $P_{\rm
b}$--$m_2$ relation is PSR~J1713+0747 \citep{sns+05}, and the measured
companion mass is in fact slightly lower than the \citet{ts99a}
prediction.  Mass measurements or constraints in more systems, by
timing or by optical spectroscopy of the WD companions, will be needed
to confirm or refute our present conclusions.  The new systems will
likely lend themselves to observations of geometrical effects such
as the change in apparent semi-major axis due to motion of the pulsar
and/or the Earth \citep{kop95,kop96}, and J1910+1256 in particular has
sufficient timing precision that it may be possible to measure Shapiro
delay in this system.

\subsection{Equivalence Principle Violations\label{sec:gr}}

The WBMSPs are the best objects for setting limits on violations of
the Strong Equivalence Principle (SEP) and the Parametrized
Post-Newtonian parameter $\alpha_3$, which describes Lorentz
invariance and momentum conservation.  These tests use the fact that
the gravitational self-energy of the NS will be much higher than that
of the WD, and therefore, if the equivalence principles are violated,
the two objects will accelerate differently in an external
gravitational field or under the self-acceleration induced by the
velocity relative to a preferred reference frame.  The net effect on
orbits that are nearly circular will be to force the eccentricity into
alignment with this acceleration vector \citep{ds91}.  The prototype
of these tests is the search for orbital polarization in the
Earth-Moon system \citep{nor68b}, which currently sets a limit on the
weak-field violation parameter $|\eta|$ of 0.001 \citep{dbf+94,wil01}.
The pulsar versions of these experiments test the strong-field limit
of SEP violation (parameter $\Delta$) and Lorentz
invariance/momentum conservation (parameter $\hat \alpha_3$), and are
thoroughly described in the literature
\citep{ds91,wex97,bd96,wex00,sta03,sns+05}.  Both parameters are
identically zero in GR, and $\hat \alpha_3$ is predicted to be zero by
most theories of gravity.

We now examine the impact of the recently discovered binary systems on
these tests.  The traditional figures of merit for choosing systems to
test $\Delta$ and $\hat \alpha_3$ are $P_{\rm b}^2/e$ and $P_{\rm
b}^2/(Pe)$, respectively.  The other requirements for $\Delta$ are
that each system must be old enough (ie have characteristic age large
enough) and must have $\dot \omega$ large enough that the longitude of
periastron can be assumed to be randomly oriented; and that each
system must have $\dot \omega$ larger than the rate of Galactic
rotation, so that the projection of the Galactic acceleration vector
onto the orbit can be considered constant \citep{ds91,wex97}.  Similar
requirements hold for $\hat \alpha_3$.  With its extraordinarily low
eccentricity, PSR~J1853+1303 is a prime candidate to help strengthen
these tests.  The last few years have seen the discovery of several
other systems with comparable or longer-period orbits, notably
PSRs~J2016+1948 \citep{naf03} and J0407+1607 \citep{lf05,lxf+05}.  We
therefore find it worthwhile to update the multi-pulsar analysis of
Wex (1997, 2000), finding much lower limits on each parameter.  In
keeping with the spirit of \citet{wex00}, we use all 21 pulsars listed
in Table~\ref{tab:allbin}, as these are all thought to have evolved
with similar extended accretion periods and therefore represent the
overall population of such objects.  Some of these systems have quite
small values of $P_{\rm b}^2/e$ but need to be included nonetheless,
as possible examples of violation.  Our calculation will find a
median-likelihood value of $|\Delta|$ for each pulsar that corresponds
to an induced eccentricity roughly comparable to its observed
eccentricity, and the combined limit fairly represents the limits
derivable from the known population.

We use the following Bayesian analysis.  For the SEP $\Delta$ test, we
are interested in finding the probability density function (pdf)
$p(|\Delta|\,|D,I)$, where $D$ represents the relevant data on the 21
pulsars (namely, their eccentricities and longitudes of periastron and
associated measurement errors) and $I$ represents prior information.
The unknown parameters for each system include the two stellar masses,
the distance $d$ to the system, and the position angle $\Omega$ of the
Line of Nodes on the sky.  Given any set of these parameters, a
``forced'' eccentricity vector $\bf {e_{\rm F}}$ may be derived for
any given value of $\Delta$, up to a sign ambiguity which amounts to
flipping the direction of the vector. This can be written \citep{ds91}:
\begin{equation}
|{\bf e}_{\rm F}| = \Delta \frac{1}{2}\frac{{\bf g}_{\perp}c^2}{FG\,(m_1+m_2) (2\pi/P_{\rm b})^2},
\label{eq:ef2}
\end{equation}
where $c$ is the speed of light, and, in general relativity, $F = 1$
and $G$ is Newton's constant.  Here ${\bf g}_{\perp}$ is the
projection of the Galactic acceleration vector onto the plane of the
orbit, and is given by \citep{ds91}:
\begin{equation}
|{\bf g}_{\perp}| = |{\bf g}|[1-(\cos i \cos\lambda + \sin i \sin\lambda\sin(\phi -\Omega))^2]^{1/2},
\label{eq:gperp}
\end{equation}
where $\phi$ is the position angle of the projection of the
gravitational acceleration vector ${\bf g}$ onto the plane of the sky,
and $\lambda$ is the angle between the line from pulsar to Earth and
${\bf g}$.  Deriving $\bf {e_{\rm F}}$ requires knowledge of the
Galactic acceleration at the pulsar position; we assume the vertical
potential given by \citet{kg89} and a flat rotation curve with
velocity of 222\, km\,s$^{-1}$.  A prediction for the observed
eccentricity ${\bf e_{\rm obs, pred}}$ is then the vector sum of ${\bf
e_{\rm F}}$ and a ``natural'' eccentricity $\bf {e_{\rm N}}$.  Thus
the magnitude of $\bf {e_{\rm N}}$ and the angle $\theta$ between $\bf
{e_{\rm N}}$ and $\bf {-e_{\rm F}}$ are additional parameters.

For any one $j$ of the 21 pulsars, we may then use Bayes' theorem to
write:
\begin{eqnarray}
 & & p(|\Delta|, i, m_2, \Omega, d, e_{\rm N}, \theta| D_j, I)\\ \nonumber
& \propto& 
p(D_j |\, |\Delta|, i, m_2,\Omega, d, e_{\rm N}, \theta, I)
p(|\Delta|, i, m_2, \Omega, d, e_{\rm N}, \theta| I).
\end{eqnarray}
\noindent We compute the integral over $e_{\rm N}$ and $\theta$ separately,
effectively calculating the marginal $p(|\Delta|, i, m_2, \Omega, d | D_j, I)$
for a particular set of parameters $i$, $m_2$, $\Omega$, and $d$.  For
each such set of parameters, the angle between ${\bf e_{\rm F}}$ and
the true ${\bf e_{\rm obs}}$, and hence the possible values of ${\bf
e_{\rm N}}$ for which the likelihood is significantly non-zero, will
almost always be very tightly constrained, due to the small
measurement errors on $e_{\rm obs}$ and $\omega$.  We therefore
determine the four points representing $3$-$\sigma$ ranges in both
$e_{\rm obs}$ and $\omega$, and use these to find minimum and maximum
values of $e_{\rm N}$ and $\theta$.  The likelihood for the set of
parameters $i$, $m_2$, $\Omega$, and $d$ is then set to 1 for values
of $e_{\rm N}$ and $\theta$ fall between their minimum and maximum
values and 0 otherwise.  Thus the integral will be roughly
proportional to $(\theta_{\rm max} - \theta_{\rm min})(\log e_{\rm N,
max} -\log e_{\rm N, min})$, assuming a uniform prior on $\theta$ and
a log prior on $e_{\rm N}$.  We set the integral to zero if $e_{\rm
N, min}> 0.05$ and use $1\times10^{-6}$ as a lower bound on possible
values of $e_{\rm N, min}$; this conservatively allows each pulsar
much more than the eccentricity ranges permitted by \citet{phi92b}.
This approximation to the likelihood is necessary as both numerical
integration over or (equivalently) Monte Carlo sampling of the full
allowed ranges of $e_{\rm N}$ and $\theta$ would be computationally
prohibitive.

\begin{figure}
  \plotone{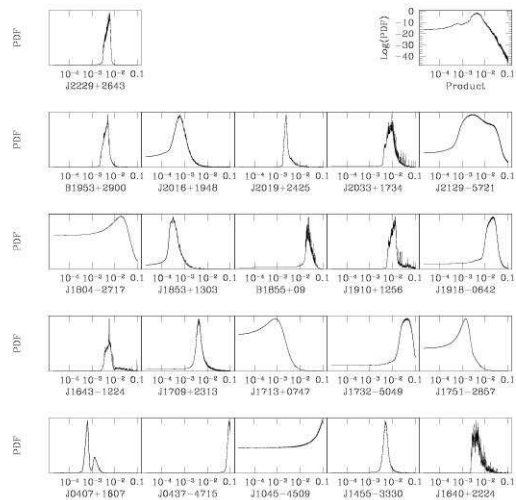} \caption{Posterior probability density
  functions (pdfs) for the test of the Strong Equivalence Principle
  (SEP). The pdf for each pulsar is shown on a linear vertical scale.
  The horizontal axis is displayed logarithmically for clarity,
  although the range $0<|\Delta|<0.1$ was sampled uniformly. The pdfs
  for some of the pulsars are noisy for those cases where the ${\bf
  e_{\rm obs}}$ and ${\bf g_{\perp}}$ vectors can be close to alignment
  for certain values of $\Omega$; these cases are difficult to model
  even with large number of trial systems, but the noise does not
  drastically affect the full pdf.  The full pdf $p(\Delta|D,I)$ is
  the normalized product of the individual-pulsar pdfs and is shown on
  a log-log scale in the uppermost right-hand panel.  \label{fig:pdf_sep}}
\end{figure}

\begin{figure}
  \plotone{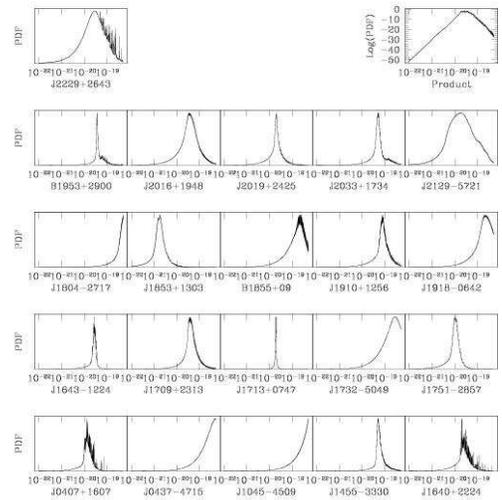} \caption{Posterior pdfs for the $\hat \alpha_3$
  test, similar to the plots in
  Figure~\ref{fig:pdf_sep}. \label{fig:pdf_a3}}
\end{figure}

For the remaining nuisance parameters $i$, $m_2$, $\Omega$, and $d$ we
perform a Monte Carlo simulation, drawing $m_2$ uniformly from twice
the \citet{ts99a} $P_{\rm b}$--$m_2$ range, $\cos i$ from a uniform
distribution between 0 and 1 (combining these two parameters and the
mass function yields a value of the pulsar mass $m_1$; we use only
those systems for which $m_1$ is between $1.0\,M_{\odot}$ and
$2.5\,M_{\odot}$, or other limits set by timing of the individual
pulsar), $\Omega$ from a uniform distribution between $0^{\circ}$ and
$360^{\circ}$ and $d$ from a Gaussian distribution about the best
\citet{cl02} distance, assuming a distance uncertainty of 25\%, or
from a Gaussian distribution in parallax, where measured.  For
PSR~J1713+0747, we restrict the parameters $i$, $\Omega$, $m_1$ and
$m_2$ to the region constrained by the recent measurement of the
orientation of the orbit
\citep{sns+05}, while for PSR~J0437$-$4715 we assume Gaussian
distributions about the parameters given in \citet{vbb+01}.  We repeat
this procedure for values of $\Delta$ from 0.00002 to 0.1, in steps of
0.00002, then normalize; this results in the posterior pdf
$p(|\Delta|\,|D_j,I)$ for each pulsar $j$.  For PSR~J1713+0747 alone, the
resulting 95\% confidence limit on $|\Delta|$ is about 0.0158, similar
to the value derived in
\citet{sns+05}.  The pulsar data sets are independent, and thus we
multiply the pdfs to derive $p(|\Delta|\,|D,I)$.  From this, we derive a
95\% confidence upper limit on $|\Delta|$ of $5.6\times10^{-3}$.
Figure~\ref{fig:pdf_sep} shows the pdf curves for this test.  We note
that while a logarithmic prior on $|\Delta|$ would result in an upper
limit a few orders of magnitude smaller, we have chosen a uniform
prior on $|\Delta|$ in order to be as conservative as possible and
more consistent with previous work.

For the $\hat \alpha_3$ test, we proceed in a similar fashion.  Here
the forced eccentricity is \citep{bd96}:
\begin{equation}
|{\bf e}_F| = \hat \alpha_3\frac{c_{\rm p}|{\bf w}|}{24\pi}\frac{P_{\rm b}^2}{P}\frac{c^2}{G(m_1+m_2)}\sin\beta
\label{eq:efalpha3}
\end{equation}
where $c_{\rm p}$ is the gravitational self-energy fraction of
``compactness'' of the pulsar, approximated by $0.21 m_1$
\citep{de92,bd96}, and $\beta$ is the (unknown) angle between the pulsar's
absolute velocity ${\bf w}$ (relative to the reference frame of the Cosmic
Microwave Background) and its spin vector.  For this test,
we also need the 3-dimensional velocity of the system.  Where proper
motion measurements are available, we draw from Gaussian distributions
for the proper motion to get the transverse velocities; in other
cases, and always for the unknown radial velocities, we draw from
Gaussian distributions in each dimension centered on the Galactic
rotational velocity vector at the pulsar location and with widths of
80\, km\,s$^{-1}$ \citep{lml+98}.  We sample uniform steps of $\hat
\alpha_3$ ranging from $1\times 10^{-22}$ to $5\times 10^{-19}$. We
find a 95\% confidence upper limit on $|\hat \alpha_3|$ of
$4.0\times10^{-20}$.  Figure~\ref{fig:pdf_a3} shows the pdf curves for
this test.

The 95\% confidence limits we derive of $5.6\times 10^{-3}$ for
$|\Delta|$ and $4.0\times10^{-20}$ for $|\hat \alpha_3|$ are
considerably better than previous limits of $9\times10^{-3}$ and
$1.5\times 10^{-19}$, respectively \citep{wex00}, while still taking
into account the contribution from all pulsars with similar
evolutionary histories.  The SEP test appears weaker than the best
solar-system tests of $|\eta| < 0.001$ \citep{dbf+94,wil01} but
pulsars test the strong-field regime inaccessible to the solar-system
measurements and are therefore qualitatively different.  The $|\hat
\alpha_3|$ test is nearly thirteen orders of magnitude better than
tests derived from the perihelion shifts of Earth and Mercury
\citep{wil93}, and again tests the strong-field regime of gravity.

\section{Conclusions} 
\label{sec:conc}

Of the three new WBMSP systems presented here, at least two can be
timed at the microsecond level with current instrumentation at large
telescopes.  These objects thus show promise for measurement of
geometrical and/or relativistic timing phenomena in future, giving us
an idea of the system inclination angles and perhaps the masses of the
objects.  This information will be extremely useful in evaluating the
validity of the $P_{\rm b}$--$m_2$ relations for estimations of
companion masses.  All three systems should provide proper motion
measurements within a few years; these will add to our understanding
of millisecond pulsar velocities throughout their population.
Finally, in combination with the other low-mass circular-orbit systems
discovered in recent years, the new pulsars set firmer limits on
violations of relativistic equivalence principles in the strong-field
regime of $5.6\times 10^{-3}$ for $|\Delta|$ and $4.0\times10^{-20}$
for $|\hat \alpha_3|$.  A better understanding of the low-mass
population as a whole will be necessary for further improvement of
these tests.

\section*{Acknowledgments}
The Parkes radio telescope is part of the Australia Telescope which is
funded by the Commonwealth of Australia for operation as a National
Facility managed by CSIRO.  The Arecibo Observatory, a facility of the
National Astronomy and Ionosphere Center, is operated by Cornell
University under a cooperative agreement with the National Science
Foundation.  IHS holds an NSERC UFA and is supported by a Discovery
Grant. DRL is a University Research Fellow funded by the Royal
Society.  FC is supported in part by NASA grant NNG05GA09G.  NDA, AP
and MB received support from the Italian Ministry of University and
Research (MIUR) under the national program {\it Cofin 2003}.  We thank
Norbert Wex for helpful discussions and David Nice and Jeff Hagen for
much effort in debugging the WAPP folded-profile acquisition modes.



\end{document}